\documentclass[prc,twocolumn,showpacs,superscriptaddress,floatfix]{revtex4}
\usepackage{graphicx,amssymb,amsmath}

\newlength{\ldag}
\newcommand{\cra}{a^\dagger}
\newcommand{\ana}{{a^{\phantom\dagger}\hspace{-\ldag}}}

\newcommand{\crb}{b^\dagger}
\newcommand{\anb}{{b^{\phantom\dagger}\hspace{-\ldag}}}

\newcommand{\crs}{s^\dagger}
\newcommand{\ans}{{s^{\phantom\dagger}\hspace{-\ldag}}}

\newcommand{\crd}{d^\dagger}
\newcommand{\andm}{{\tilde d^{\phantom\dagger}\hspace{-\ldag}}}

\newcommand{\crt}{t^\dagger}
\newcommand{\ant}{{t^{\phantom\dagger}\hspace{-\ldag}}}

\begin{document}
\title{A scalar two-level boson model to study the IBM phase diagram in the Casten triangle}

\author{Julien Vidal}
\affiliation{Laboratoire de Physique Th\'eorique de la Mati\`ere Condens\'ee, CNRS UMR 7600,
Universit\'e Pierre et Marie Curie, 4 Place Jussieu, 75252 Paris Cedex 05, France}

\author{Jos\'e M. Arias}
\affiliation{Departamento de F\'{\i}sica At\'omica, Molecular y
Nuclear, Facultad de F\'{\i}sica, Universidad de Sevilla,
Apartado~1065, 41080 Sevilla, Spain}

\author{Jorge Dukelsky}
\affiliation{Instituto de Estructura de la Materia, CSIC, Serrano
123, 28006 Madrid, Spain}

\author{Jos\'e Enrique Garc\'{\i}a-Ramos}
\affiliation{Departamento de F\'{\i}sica Aplicada, Universidad de
Huelva, 21071 Huelva, Spain}

\begin{abstract}
We introduce a simple two-level boson model with the same energy
surface as the Q-consistent Interacting Boson Model Hamiltonian. The model can be
diagonalized for large number of bosons and the results used to
check analytical finite-size corrections to the energy gap and the
order parameter in the critical region.
\end{abstract}


\maketitle

\section{Introduction}

The study of quantum phase transitions  is a hot topic covering
different branches of quantum many-body physics. In recent years,
the study of low-dimensional lattice models has been triggered by
the increasing interest on the development of a quantum computer.
Moreover, concepts of quantum information have been used to
characterize quantum critical phenomena. On a different respect,
there is a revival of the study of structural changes in
finite-size systems where precursors of the transitions could be
observed \cite{Iachello04}. In atomic nuclei, the Interacting Boson
Model (IBM) \cite{Iachello87} provides a framework simple but
still detailed in which first and second order phase transitions can be
studied. In fact, the transition from the spherical $U(5)$
dynamical symmetry to the deformed $\gamma$-unstable $O(6)$
dynamical symmetry has been extensively studied in the last few
years \cite{Jolie02,ADG03,TuRo05}. The Hamiltonian describing this
transition is a repulsive boson pairing Hamiltonian that has the
particularity of being exactly solvable allowing the study of very
large systems. The boson pairing model was first solved  by
Richardson \cite{Richardson68} and more recently applied to the
IBM \cite{Pan98,Duke2001a, Arias03}. However, this is the only
integrable line in the phase diagram of the IBM often described
with the Q-Consistent Hamiltonian (QCH) \cite{CQIBM} and
pictorially represented by the Casten triangle. This integrable
line corresponds to one side of the triangle. Therefore, the
interior area of the triangle as well as the other two sides are
out of reach of large scale numerical studies. Numerical
calculations in these cases are then limited by the number of bosons that
standard IBM codes can manage, which are of the order of $10^2$
bosons. However, strictly speaking quantum phase transitions only
occur in macroscopic systems. Thus, it is of great interest to be
able to extend previous analysis in the Casten triangle to a
significant larger number of bosons.

In this paper, we introduce a simple two-level boson model
depending on two control parameters, that leads to the same energy
surface as the Q-consistent IBM Hamiltonian when the $\gamma$ degree
of freedom is frozen at $\gamma=0$. In other words, both energy
surfaces coincide if axial symmetry is imposed in the IBM
treatment. This model can be solved for very large boson numbers
and it allows us to investigate the properties of the first-order
phase transition from spherical to deformed axial shapes in the IBM. The
two-level boson model can be viewed as a generalized Lipkin model
with parity breaking terms, as a two coupled large spin system
\cite{GaChud2000} or as a model to study tunnelling dynamics
between two minima \cite{Instanton}.

In Sect. II, after a brief presentation of the model, we derive its
energy surface. Next, in Sect. III 
we compute the finite-size corrections up to
order $(1/N)^1$ in the spherical phase which allows us to discuss
the finite-size behaviour at the critical point for different
quantities such as the ground state energy, the gap, and the order
parameter. Finally, Sect. IV is for summarizing.

\section{The model}

Let us now introduce the two-level boson model whose Hamiltonian
is

\begin{equation}
  \label{eq:hamiltonian}
  H=x~  n_t-\frac{1-x}{N} ~ Q^{y} Q^{y},
\end{equation}
where the operators $n_t$ and $Q^{y}$ are defined as
\begin{equation}
  \label{eq:Qdef}
 n_t= \crt \ant, \hspace{0.5cm} Q^{y}=\crs \ant+\crt \ans+ y ~  \crt \ant
\end{equation}
in terms of two species of scalar bosons $s$ and $t$, $x$ and $y$
being two independent control parameters.
The total number of bosons $N=n_s + n_t$ is a conserved quantity.
 The connection between the Hamiltonian
(\ref{eq:hamiltonian}) and a generalized Lipkin model or a
two-spin model can be obtained by making the inverse of the
Schwinger transformation $K^{+}=\crt \ans$, $K^{-}=\crs \ant$, and
$K^{0}=\frac{1}{2}(\crt \ant-\crs \ans)$.

We have deliberatively written the two-level boson Hamiltonian in
the form (\ref{eq:hamiltonian}) to resemble QCH, where $s$ and $t$
play the role of the $s$ and $d_{\mu}$ bosons of the IBM
respectively. Clearly, the difference resides on the quadrupolar
character of the $d_{\mu}$ boson leading to the $U(6)$ algebra of
the IBM, while our model is described by a $U(2)$ algebra. Despite
of this important difference, the $st$ model captures the main
characteristics of the phase diagram of IBM as we will now
show.

The phase diagram of model (\ref{eq:hamiltonian}) can be easily
obtained in the coherent state approach. Therefore, we introduce a
variational state of the form

\begin{equation}
\label{Cohe}
\left\vert N,\beta \right\rangle
=e^{\sqrt{\frac{N}{1+\beta ^{2}}}\left( s^{\dagger }+\beta
t^{\dagger }\right) }\left\vert 0\right\rangle
\end{equation}
where $|0\rangle$ denotes the boson vacuum. The corresponding
energy surface as a function of the variational parameter $\beta$, in
the large $N$ limit, is given by:
%
%
\begin{eqnarray}
\label{eq:ES}
E(N,\beta)&=&\langle N,\beta  | H | N,\beta \nonumber \rangle\\
&=& N {\beta^2 \over (1+\beta^2)^2} \Big\{ 5x-4+4 \beta  y (x-1)+ \nonumber \\
&&\beta^2\big[x+y^2(x-1) \big] \Big\} .
\end{eqnarray}

Minimization of the energy (\ref{eq:ES}) with respect to $\beta$,
for given values of the control parameters $x$ and $y$, gives the
equilibrium value $\beta_0$ defining the phase of the system.
$\beta_0=0$ corresponds to the symmetric phase, and $\beta_0\neq
0$ to the broken symmetry phase. Here, we restrict the study to
the range $x\in [0,1]$ and $y>0$  which implies $\beta_0 \geq 0$.
However, the treatment can be easily extended to the whole
parameter space.

\begin{figure}[h]
  \centering
  \includegraphics[width=6.5cm]{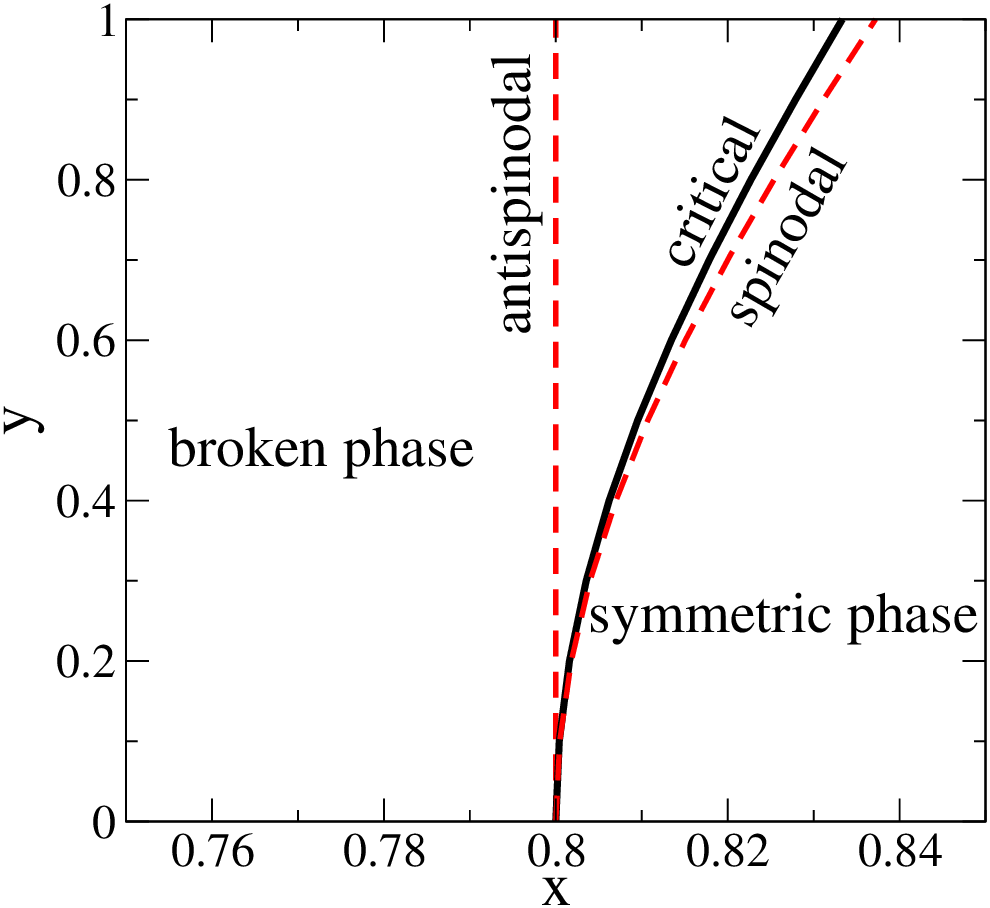}
  \caption{(Color online) Phase diagram for the Hamiltonian (\ref{eq:hamiltonian}) in
    the plane of the control parameters $(x,y)$. For $y=0$, the transition is second-order
    whereas it is first-order otherwise.}
  \label{phaseD}
\end{figure}

The schematic phase diagram of the $st$ model is presented in Fig.
\ref{phaseD} in the region of interest around the phase
transition. In addition to the critical line displayed by the
solid curve, we show the spinodal and antispinodal lines where the
second minimum in the energy surface (\ref{eq:ES}) starts to
appear indicating a region in which both phases coexist.

\begin{figure}[h]
  \centering
  \includegraphics[width=8.0cm]{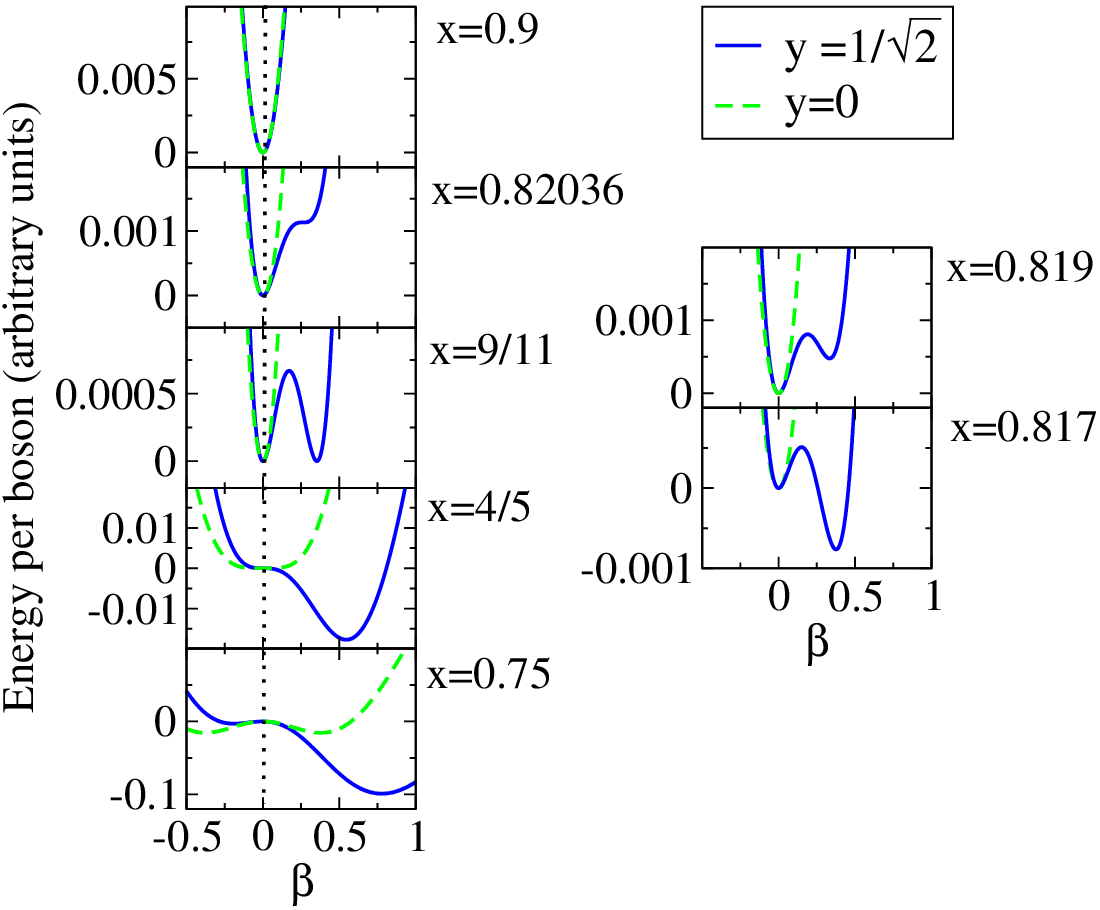}
  \caption{(Color online) Energy surfaces for the Hamiltonian
    (\ref{eq:hamiltonian}) in the large $N$ limit (\ref{eq:ES}) for
    $y=1/\sqrt{2}$ (full line) and $y=0$ (broken line) and for
    different $x$-values as a function of the deformation parameter
    $\beta$. The limits of the coexistence region are 
  shown in the panels $x=0.82036...$ (spinodal) and $x=4/5$ (antispinodal). The
critical point is represented in the middle (left) panel ($x=9/11$). The two panels on the right show two cases in the
coexistence region.}
  \label{fig1}
\end{figure}

For $y=0$, there is an isolated point of second order phase
transition as a function of $x$. Spinodal, antispinodal and
critical point coincide at the critical value $x_c=4/5$. For $y
\neq 0$ the phase transition changes its character to first order
as can be seen in Fig. \ref{fig1} for $y=1/\sqrt{2}$.
Schematically, for $x=1$, the system is in the symmetric phase
since the energy surface has a unique minimum at $\beta=0$. When
$x$ decreases, one reaches the spinodal point ($x=0.82036$ for
$y=1/\sqrt{2}$) where a second local (nonsymmetric) minimum ($\beta
\neq 0$) arises. This nonsymmetric minimum  competes with the symmetric one till both attain the same depth at the critical point $x_c=(4+y^2)/(5+y^2)$. Beyond this
value, the symmetric minimum at $\beta=0$ becomes a local minimum
till $x=4/5$ where it becomes unstable (anti-spinodal point). We
show in Fig. \ref{fig1}, (left panels), a sketch of this evolution for the
special cases $y=0,1/\sqrt{2}$. The right panels in Fig. \ref{fig1} show
two cases in the coexistence region.

\begin{figure}[h]
  \centering
  \includegraphics[width=7.0cm]{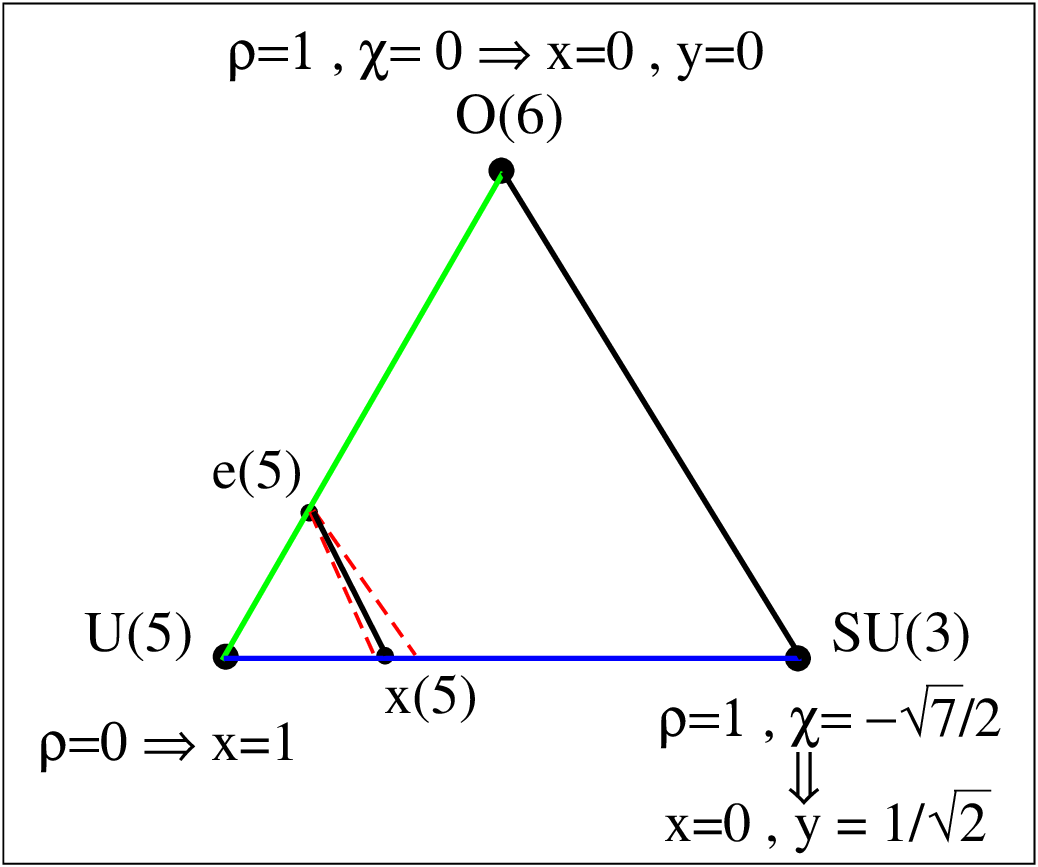}
  \caption{(Color online) Usual representation of the IBM phase diagram as a Casten
    triangle. The three dynamical symmetries are located in the
    vertices. Critical points $e(5)$ (second order) and $x(5)$ (first
    order) with the coexistence region are also included. 
Variation of the control parameters, either in
    IBM or in the present model, allows to explore the whole model space
    represented by the triangle.}
  \label{figCasten}
\end{figure}

It is worth mentioning that this scheme is exactly the same as the
one obtained in the IBM of nuclear structure. In fact the value
$y=1/\sqrt{2}$ is equivalent to $\chi=-\sqrt{7}/2$ describing the
$U(5)$ to $SU(3)$ side of the Casten triangle \cite{Casten} as we are
showing now. The energy (\ref{eq:ES})
coincides with that of the Q-consistent IBM-1 Hamiltonian in the
thermodynamic limit (large $N$). 
This correspondance is readily established if we write the
Q-consistent IBM-1 Hamiltonian  
\begin{equation}
  \label{eq:hamiltonianIBM}
  H_{IBM}=x~  n_d-\frac{1-x}{N} ~ Q^{\chi}\cdot Q^{\chi},
\end{equation}
where $n_d$ is the $d-$boson number operator and $Q^{\chi}$ is defined as
\begin{equation}
  \label{eq:QdefIBM}
Q^{\chi}=\crs \andm+\crd \ans+ \chi ~  \left(\crd \times \andm\right)^{(2)},
\end{equation}
with $\chi$ the structure parameter of the quadrupole operator of the IBM.
For the Hamiltonian (\ref{eq:hamiltonianIBM}) the energy surface in
the large $N$ limit reads (see for instance Eq. (6) in
Ref. \cite{Cejnar00} taking the large $N$ limit and $\gamma=0$, this
last condition is not a restriction since it is known that the IBM-1
including up to two--body interactions does not produce triaxiality)
\begin{eqnarray}
\label{eq:ESIBM}
E_{IBM}(N,\beta)&=&\langle N,\beta  | H_{IBM} | N,\beta \nonumber \rangle\\
&=& N {\beta^2 \over (1+\beta^2)^2} \Big\{ 5x-4-4 \sqrt{\frac{2}{7}} \beta \chi (x-1)+ \nonumber \\
&&\beta^2\big[x+\frac{2}{7}~ \chi^2 (x-1) \big] \Big\} .
\end{eqnarray}
The equivalence between (\ref{eq:ES}) and (\ref{eq:ESIBM}) is readily established 
 in the thermodynamical limit setting $y=-\sqrt{2/7} \chi$. 
 This correspondence can also be obtained from (\ref{eq:QdefIBM}) if one is restricted to the spherical $\mu=0$ component of the $d$-bosons. In that case
\begin{equation}
  \label{eq:QdefIBM2}
Q=\crs \andm_0+\crd_0 \ans+ \chi ~ \langle 2 0 2 0|2 0\rangle \crd_0 \andm_0,
\end{equation}
with the Clebsch-Gordan coefficient  $\langle 2 0 2 0|2 0\rangle=-\sqrt{2/7}$. 
 It should be clear that this equivalence
is only valid in the thermodynamical limit and restricted to the scalar
excitations.  
Differences between both models are due to the fluctuations induced by
the $\gamma$--vibration over the mean field. The present scalar model
does not include $K=2$ excitations and consequently can not be used
to study $\gamma$--excitations. But, apart from that, it
mimics the usual representation of the IBM in a Casten triangle
as shown in Fig. \ref{figCasten}, where any point in the triangle can
be described in polar coordinates $(\rho,\theta)$. The radial variable $\rho$
is related to $x$ (basically $\rho=1-x$) and the angular variable
$\theta$ is related to $y$ ($\theta= (1-\sqrt{2} y)\pi/3$)
(or $\chi$ in the IBM). Thus, changing the control parameters ($x,y$)
(or $x,\chi$ in IBM) one can reach any point in the triangle and,
consequently analyze any trajectory in it looking for critical
phenomena. In this respect, the present model allows to perform
numerical calculations within the Casten triangle for large $N$
values, which are not posible with the usual IBM codes and are needed
to check finite-size corrections of different observables as we will
show in the next section.

\begin{figure}[h]
  \centering
  \includegraphics[width=6.5cm]{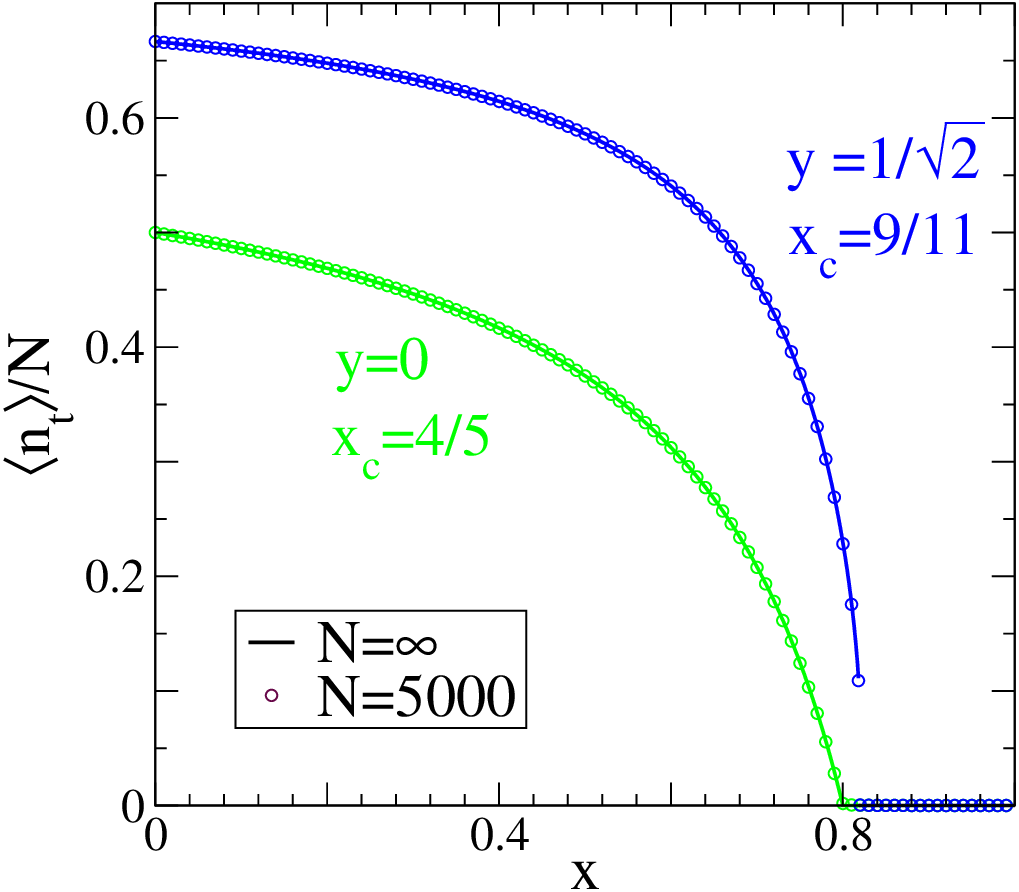}
  \caption{(Color online) Variation of the order parameter as a function of $x$ near
    the critical point. Full line are the analytical results for
    $N=\infty$ and circles are numerical results for $N=5000$. The
    jump is clearly observed for $y\neq 0$.} 
\label{fig2}
\end{figure}

We would like to emphasize here that the $st$ boson model has a
natural order parameter, the expectation value of the $t$-boson
number operator $n_t$. This is not the case in the spin
representation of the model \cite{GaChud2000}. On this regard, the
Schwinger mapping could be used to translate the number operator
$n_t$ into spin operators suggesting an order parameter for the
latter model. In the coherent state representation 
($N\rightarrow \infty$) the expectation value of the $t$-boson number
operator is simply given by $\langle n_t \rangle/N = {\beta^2 /
  (1+\beta^2)}$. For $x=1$, this order parameter vanishes whereas for
$x=0$ it is given by $1-\frac{2}{4 + y^2 + \sqrt{y^2(4 + y^2)}}$.
At the transition point $x_c$, both minima lead to the same energy
and the jump of the order parameter is $y^2/(4+y^2)$ which
vanishes, as expected, for $y=0$. We have displayed in Fig.
\ref{fig2} the behaviour of the order parameter as a function of
$x$ for two different values of $y$.

\section{Finite size corrections: beyond mean field }

For $y=0$, the $st$ model corresponds, up to some redefinitions of
the parameters,  to the LMG model which has been recently shown to
display nontrivial finite-size scaling properties at the
transition point \cite{Dusuel04_3,Dusuel05_2,Leyvraz05}. As
explained above, the transition for $y \neq 0$ is of first order
and it is thus natural to expect some qualitative changes for the
finite-size corrections. To analyze them, we follow the same route
as those detailed in Ref. \cite{Dusuel05_2} for  $y = 0$.

For sake of clarity, we briefly sketch now this approach in the
symmetric phase and refer to Ref. \cite{Vidal06_1} for more
details. For $x>x_c$, we map the one-body operators using the
Holstein-Primakoff boson expansion \cite{Holstein40} onto  a
single $b$-boson

%
%
\begin{eqnarray}
\label{eq:def1}
\crt \ant&=& \crb \anb,\\
\crt \ans &=& N^{1/2} \crb (1-n_b/N)^{1/2}=(\crs \ant)^\dag,\\
\crs \ans &=& N-n_b.
\label{eq:def3}
\end{eqnarray}
%
%
with $\langle n_b \rangle/N \ll 1$.
At order $p$, in the symmetric phase, the Hamiltonian schematically reads:
%
%
\begin{equation}
H = \sum_{i=0}^{p} {H_i  \over  N^{i}} +O(1/N^{p+1}).
\label{eq:hamilHP}
\end{equation}
%
%
We make use of the canonical transformation method which allows to
diagonalize exactly the Hamiltonian order by order. At order
$(1/N)^0$, $H$ is simply a quadratic form and is thus simply
diagonalized via a Bogoliubov transform. As explained in  Ref.
\cite{Vidal05}, diagonalizing $H$ beyond this order only requires
to solve a linear set of algebraic equations. More concretely, let
us introduce the $a$-boson through:
%
%
\begin{equation}
\crb=  \sum_{j= 0}^{p} {1 \over N^j} \sum_{k,l} \alpha_{k,l}^{(j)}
{\cra}^{k} {\cra}^{l} ,
\label{eq:cano}
\end{equation}
%
%
where the $\alpha_{k,l}^{(j)}$ are coefficients to be determined such
that $H$ expanded at order $1/N^p$ is a polynomial function in
$n_a$. Further, one also has to impose the bosonic commutation rules
which reads $[\anb,\crb]=1$.  At each order, the content in $\cra$ and
$\ana$ is given by a simple power counting analysis.
For $p=0$, this transformation is nothing but the Bogoliubov
transformation $\crb=\alpha_{1,0}^{(0)} \cra+\alpha_{0,1}^{(0)}  \ana$
and we thus have two coefficients to determine which are solutions of
quadratic equations.
At order $p=1$, one must include linear and cubic terms and we have
six coefficients $\alpha_{k,l}^{(1)}$ to find which are now solutions
of a linear set of equations. One then gets, at order $1/N$, the
following diagonal form of the Hamiltonian:

\begin{widetext}
%
%
\begin{eqnarray}
  H &=&{1 \over 2} \left[-x+\Xi(x)^{1/2} \right]+ \nonumber \\
&&  {x^2(x-1) \over 2 N} \left[
{-8+26x-20x^2+y^2(-2+6x -3x^2) \over \Xi(x)^2} +
x {32-80x +50x^2+y^2(8-14x+5x^2)\over \Xi(x)^{5/2}}
  \right]+ \nonumber \\
&&n_a \left\{
\Xi(x)^{1/2} +  {x^2(x-1) \over N}
\left[
{-8+10x+y^2(-2+x) \over \Xi(x)^{3/2}}+ {-16+52x-40x^2+y^2(-4+16x-10x^2) \over \Xi(x)^{2}}
\right]
\right\} + \nonumber \\
&&
:n_a^2: {x^2(x-1) \over N} {-8+26 x- 20 x^2+ y^2(-2+8x-5x^2)  \over \Xi(x)^{2}}+O(1/N^{2}),
\end{eqnarray}
%
%
\end{widetext}
with $\Xi(x)=x(5x-4)$. As can be readily seen, the ground state energy
as well as the gap are regular functions provided $x>4/5$ which is the
value of the critical point for $y=0$. For nonvanishing $y$, it means
that for $x \geq x_c>4/5$, there is no divergence in these
quantities. One thus naively expect some $1/N$ corrections from this
side of the transition.

\begin{figure}[t]
  \centering
  \includegraphics[width=6.5cm]{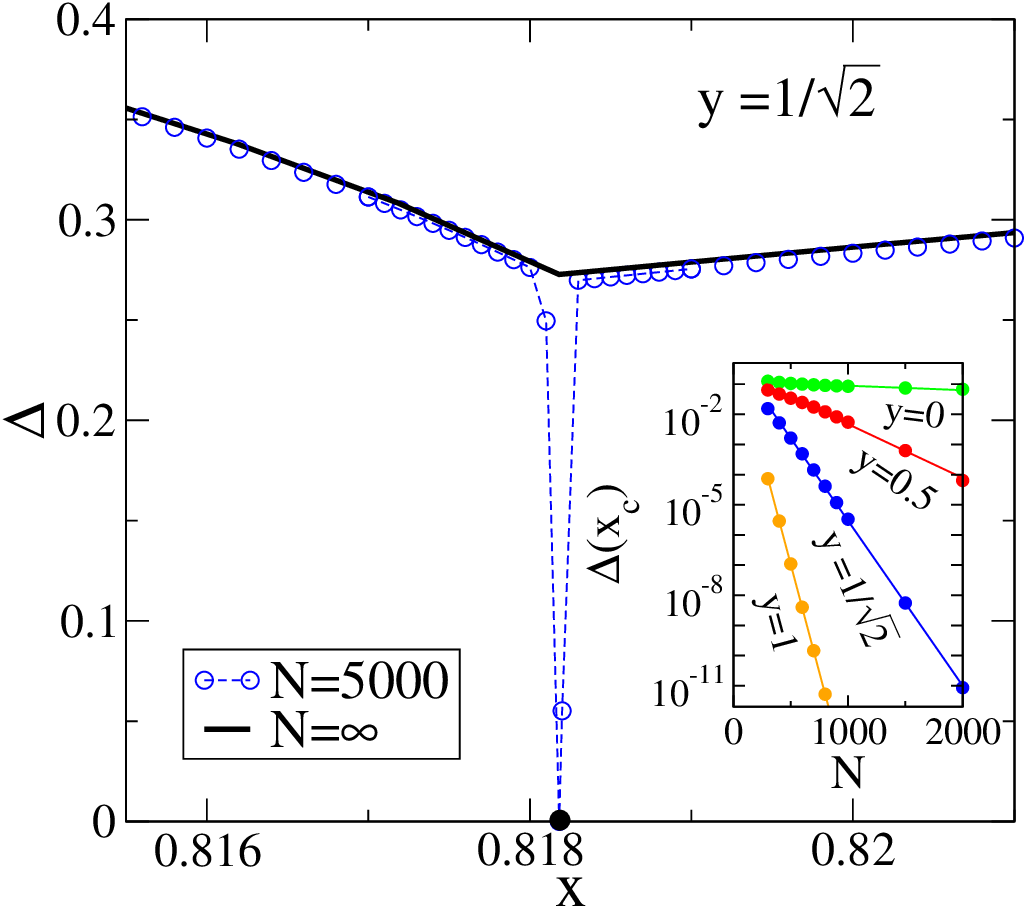}
  \caption{(Color online) Behaviour of the gap as a function of
    $x$ near the critical point for $y=1/\sqrt{2}$. The full bold line
    and the full dot at $x_c=9/11$ correspond to the analytical results.
    Open dots are the numerical calculation for $N=5000$. In the inset, the
    exponential decrease of the gap at the critical point is shown for
  several values of $y$.}
  \label{figaver1}
\end{figure}

\begin{figure}[t]
  \centering
  \includegraphics[width=6.5cm]{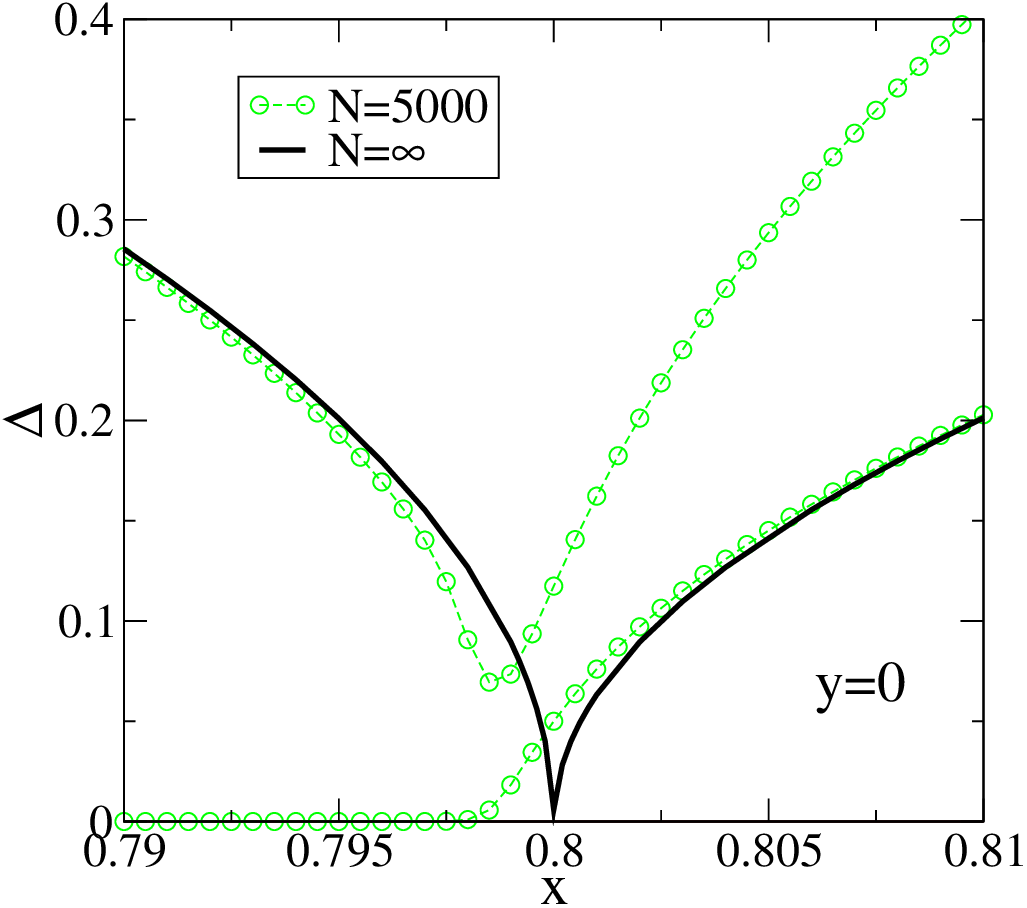}
  \caption{(Color online) Same as Fig. \ref{figaver1} but for $y=0$. The full line is the analytical result for the gap, while the open dots correspond to the first two excited states for
  $N=5000$.}
  \label{figaver2}
\end{figure}

The same approach can obviously be applied in the broken phase to
diagonalize $H$ at the same order. It then requires the exact
knowledge of $\beta_0$ which is solution of the cubic equation
$\partial E(N,\beta) / \partial \beta=0$. If, for arbitrary $0 <x
< x_c$, it is a complicated expression of $x$ and  $y$, the
nonvanishing solution for $x=x_c$ simply reads $\beta_0= y/2$ and
thus allows for simplifications. The main result of this study is
that the Hamiltonian expanded at order $1/N$ is exactly the same
at the critical point $x=x_c$ for $\beta_0=0$ and $\beta_0= y/2$.
Of course, at order $N$, one finds $H=0$ which simply confirms the
existence of a first order quantum phase transition. The most interesting point is that
the finite-size corrections are also the same for both values of
$\beta_0$. Physically, it means that for $x=x_c^\pm $ the gap is
$\Delta=\Xi(x_c)^{1/2} +O(1/N)$ but for $x=x_c$, the gap must
vanish in the large $N$ limit.
In Fig. \ref{figaver1} we show the behaviour of the gap in the region around the critical point,
$x_c=9/11$, for $y=1/\sqrt{2}$.  As can be seen gaps in the symmetric
and the broken phases are indeed equal at $x=x_c^\pm$. However,
exactly at the critical point both 
states are degenerated giving rise to a zero gap as indicated by
the black dot at $x_c=9/11$. In the inset, we show the behaviour of the
gap at the critical point as a function of the boson number $N$,
confirming the predicted 
exponential  $\Delta \sim e^{-A N}$. The exponent $A$ is a
function of $y$ that vanishes in the limit  $y=0$ where we expect
$\Delta \sim N^{-1/3}$ \cite{Dusuel05_3}. This exponential decay
clearly indicates that the $1/N$ expansions from the symmetric
phase ($\beta_0=0$) and from the broken one ($\beta_0= y/2$) are
the same at all orders since any difference would imply a power-law
decrease. As will be discussed in a forthcoming publication for
the standard IBM model, we emphasize that this matching of the
$1/N$ expansions is not a rule but rather an exception
\cite{Vidal06_1}.
In Fig. \ref{figaver2} the gap for $y=0$ is shown in the region around
the critical point. The full line corresponds to  the analytical
expression and vanishes at 
the critical point. In this case, there are no parity breaking terms so results
discussed in Ref. \cite{Dusuel05_2} are recovered. For $x>4/5$, the
system is in the symmetric phase with positive parity. The gap is
related to the energy of the first excited state 
corresponding to a one phonon state with negative parity. The
second excited state is a two phonon state with positive parity.
The  excitation energies for these two states for $N=5000$ are
plotted in Fig. \ref{figaver2} as circles joined by a (green) line. 
For $x<4/5$ the system is in the broken phase. In the limit $x=0$ there are two sets of
states of different ``parity'' degenerated ($\pm$ parity doublets) as
displayed by the lowest (green) circle-line in Fig. \ref{figaver2}. 
In this phase the nonvanishing gap is
related to the excitation energy of a different phonon which
corresponds to the upper (green) circle-line.
It is equivalent to the $\beta$ excitation in IBM (there are no
$\gamma$ excitation in this scalar model). In the limit $N\rightarrow
\infty$ both gaps (from the spherical and from the broken sides) go to
zero at the critical point, although they remain finite at finite $N$
as clearly seen in Fig. \ref{figaver2}. 

\section{Summary and conclusions}

To summarize, we have introduced a two-level model in terms of two
scalar bosons $s$ and $t$,  whose phase diagram in the
thermodynamical limit is the same as the IBM phase diagram when
described by the QCH. We have exactly diagonalized the Hamiltonian in the symmetric phase up to order $(1/N)^1$ and obtained the finite-size corrections  for the ground state energy, the gap and the order parameter. These corrections were tested against
numerical results for large systems. Though the phase diagram of this $st$ model and the QCH coincide in the large $N$ limit, beyond the mean-field level
the QCH may have different properties due to the dynamics of fluctuations and correlations in the full ($\beta$,$\gamma$) space \cite{Cejnar00,Rowe04_2,Rowe05}.
In particular, while our $st$ model is fully integrable \cite{Benet03} the IBM is
chaotic for $\chi\neq 0$ \cite{Alhassid90} with the exception of the
$SU(3)$ symmetry limit. Extension of the formalism presented in this
paper to the IBM will be the subject of a forthcoming publication \cite{Vidal06_1}.

\section{Acknowledgments}

This work has been partially supported by the Spanish Ministerio
de Educaci\'on y Ciencia
and by the European regional development fund (FEDER) under
projects number BFM2003-05316-C02-02, FIS2005-01105 and FPA2003-05958.



\end{document}